\documentstyle[12pt]{article}
\newcommand{\be}{\begin{equation}}
\newcommand{\ee}{\end{equation}}
\newcommand{\bea}{\begin{eqnarray}}
\newcommand{\eea}{\end{eqnarray}}
\newcommand{\ba}{\begin{array}}
\newcommand{\ea}{\end{array}}     

\newcommand{\half}{\frac{1}{2}}

\begin{document}
\begin{titlepage}

\hspace{9.5cm}{IFT-P.057/97}

\vspace{.5cm}
\begin{center}
\LARGE

{\sc First Massive State of the Superstring in Superspace}

\vspace{.5cm}
\large

Nathan Berkovits\footnote{e-mail: nberkovi@power.ift.unesp.br} \\ and \\
Marcelo M. Leite \footnote{e-mail: mmleite@power.ift.unesp.br}
\vspace{.5cm}

{\it Instituto de F\'\i sica Te\'orica, Universidade Estadual 
Paulista} \\ {\it Rua Pamplona 145, 01405-900, S\~ao Paulo, SP, Brasil}

\vspace{.5cm}
September, 1997
\end{center}
\vspace{1cm}

\begin{abstract}
Using the manifestly spacetime supersymmetric description of the
four-dimensional open superstring, we construct the vertex operator 
in superspace for the first massive state. This construction provides
an N=1 D=4 superspace representation of the massive spin-2 multiplet. 
\end{abstract}
\end{titlepage}
\newpage
\section{Introduction}

\noindent

The study of massive states using the manifestly spacetime supersymmetric
description of the superstring has at least two 
interesting aspects. Firstly, we want to understand the massive 
spin-2 multiplet in superspace, which is the first excited state of the 
superstring. For example, using open superstring field theory\cite
{ref1}, 
it should be possible
to construct a superspace action for this multiplet.  
Secondly, the vertex operator of this 
massive state can be used to perform manifestly
super-Poincar\'e invariant calculations of superstring 
scattering amplitudes\cite{ref2}.

The manifestly spacetime supersymmetric description of the superstring
can be used to 
describe any four-dimensional 
compactification of the ten-dimensional superstring which preserves 
at least $ N=1$ four-dimensional spacetime supersymmetry. \cite{ref3} 
Unlike vertex operators in either the Ramond-Neveu-Schwarz or
light-cone Green-Schwarz descriptions, vertex operators in this description
are manifestly super-Poincar\'e covariant since they are constructed
in N=1 D=4 superspace. 

The plan of the paper is as follows: In section 2, we review the construction
of the vertex operator in superspace for the ground state
of the open four-dimensional superstring, which is the massless N=1 D=4 
super-Maxwell multiplet. This vertex operator is expressed in terms
of the usual scalar superfield $V$ for a super-Maxwell prepotential. 
In section 3, we construct the vertex
operator in superspace for the first excited state of the open 
four-dimensional superstring,
which is the massive N=1 D=4 spin-2 multiplet. This vertex operator
is expressed in terms of a vector superfield $V_m$ for 
the massive spin-2 prepotential. 

\section{Review of the Manifestly Spacetime Supersymmetric Description}

\noindent

The left-moving worldsheet fields in the manifestly spacetime supersymmetric
description include five bosons
$(x^{m},\rho)$ and eight fermions 
$(\theta^{\alpha}$, 
${\bar\theta}^{\dot{\alpha}}$, 
$ p^{\alpha}$,
${\bar p}^{\dot{\alpha}})$, as well as a $c=9$ N=2 superconformal
field theory for the six-dimensional compactification manifold. 
These worldsheet fields are related to the those in the RNS
description by a field-redefinition, and reduce in light-cone gauge
to the light-cone Green-Schwarz worldsheet fields. 

The worldsheet action for the left-moving fields of the open
superstring is given by
\begin{eqnarray}
S= \frac{1}{2\pi} \int d^{2}z(\frac{1}{2}\partial x^{m}\bar\partial x_{m} -
p_{\alpha}\bar\partial\theta^{\alpha} -
\bar p_{\dot{\alpha}}\bar\partial{\bar\theta}^{\dot{\alpha}} +
\frac{1}{2} \partial \rho\bar\partial \rho ) + S_{C}
\end{eqnarray}
where $S_C$ is the action for the compactification-dependent fields. 
Note that we will suppress throughout this paper
the right-moving fields of the open
superstring, which are related in the usual way
by boundary conditions to the left-moving fields.   

The above action implies the following 
free-field OPE's as $y\to z$ :
\begin{equation}
 p_{\alpha}(y)\theta^{\beta}(z) = \frac {\delta_{\alpha}^{\beta}}
 {y-z}\;,\;\;\;\;
{\bar p}_{\dot\alpha}(y){\bar\theta}^{\dot\beta}(z) 
= \frac {\delta_{\dot\alpha}^{\dot\beta}} {y-z}.
\end{equation}

\begin{equation}
\rho(y)\rho(z) = ln(y-z)\;,\;\;\;\;
x^{m}(y)x^{n}(z) = \eta^{mn} ln|y-z|.
\end{equation}

We use the conventions 
that $\sigma^m_{\alpha\dot\alpha}$ and
${\bar\sigma}_m^{\dot\alpha\alpha}$ are the Pauli matrices 
and that the Minkowski metric tensor is 
$\eta^{mn}= diag(+1,-1,-1,-1)$. In these conventions, 
a vector is transformed to a bispinor 
according to 
$v^{m} = 
\frac{1}{2}(\bar\sigma^m)^{\dot{\alpha} \alpha}v_{\alpha\dot{\alpha}}$,
and $v_{\alpha \dot{\alpha}} = (\sigma^{m})_ {\alpha \dot{\alpha}} v_{m}$.

\noindent

In this description, the superstring possesses 
a critical N=2 superconformal invariance which is related
to the topological N=2 superconformal invariance of the RNS superstring.
In terms of the above fields, 
the $c=6$ N=2 superconformal generators $(T,G,\bar G,J)$ are:
\begin{equation}
T = \frac{1}{2}\partial x^{m}\partial x_{m} 
- p_{\alpha}\partial \theta^{\alpha} 
- \bar p_{\dot\alpha}\partial \bar\theta^{\dot\alpha}
+ \frac{1}{2}\partial\rho \partial\rho 
+ T_{C} 
\end{equation}

\begin{equation}
G = e^{i\rho}d^{\alpha}d_{\alpha} + G_{C}
\end{equation}

\begin{equation}
\bar G = e^{-i\rho}\bar{d^{\dot\alpha}}\bar{d_{\dot\alpha}} + \bar G_{C}
\end{equation}

\begin{equation}
J = -\partial\rho + J_{C},
\end{equation}
where
\begin{equation}
d_{\alpha} = p_{\alpha} 
+ \frac{i}{2} \bar\theta^{\dot\alpha}\partial x_{\alpha \dot{\alpha}} 
- \frac{1}{4}(\bar\theta)^{2}\partial\theta_{\alpha} 
+\frac{1}{8}\theta_{\alpha}\partial(\bar\theta)^{2}
\end{equation}

\begin{equation}
\bar{d_{\dot\alpha}} = \bar p_{\dot\alpha} 
+ \frac{i}{2}\theta^{\alpha}\partial x_{\alpha \dot{\alpha}} 
- \frac{1}{4}(\theta)^{2}\partial\bar\theta_{\dot\alpha}
+\frac{1}{8}\bar\theta_{\dot\alpha}\partial(\theta)^{2}
\end{equation}

\begin{equation}
\Pi_{\alpha \dot{\alpha}}= \partial x_{{\alpha \dot{\alpha}}} 
+ i\theta_\alpha\partial{\bar\theta_{\dot\alpha}} 
+ i\bar\theta_{\dot\alpha}\partial\theta_{\alpha},
\end{equation}
are spacetime supersymmetric combinations of the
worldsheet fields and $[T_C,G_C,\bar G_C,J_C]$
are the $c=9$ N=2 superconformal generators of the
compactification dependent fields. 
In this paper, 
the N=2 superconformal generators are untwisted so
$G$ and $\bar G$ have conformal weight $3\over 2$ while
$e^{\pm i\rho}$ has conformal weight $-\half$. 

Note that 
$[T_C,G_C,\bar G_C,J_C]$ contain no 
singular OPE's 
with the compactification independent fields. 
Furthermore,the OPE $d_{\alpha}(y)d_{\beta}(z)$ is regular and
\begin{equation}
d_{\alpha}(y) \bar d_{\dot\alpha}(z)\to i\frac{\Pi_{\alpha
\dot{\alpha}}}{y-z}
\;,\;\;\;\; 
d_{\alpha}(y)\Pi_{\beta \dot{\beta}}(z)\to
\frac{-2i\epsilon_{\alpha\beta}\partial \bar\theta_{\dot{\beta}}}{y-z}.
\end{equation}

\noindent

Vertex operators $\Phi$ are physical if they satisfy the
N=2 super-Virasoro conditions: 
\begin{equation}
L_{n}|\Phi\rangle = 0 ,   n\geq0
\end{equation}
\begin{equation}
G_{r}|\Phi\rangle = 0,   r>0
\end{equation}
\begin{equation}
\bar G_{r}|\Phi\rangle = 0,   r>0
\end{equation}
\begin{equation}
J_{n}|\Phi\rangle = 0 ,   n\geq0.
\end{equation}
Note that $L_0|\Phi \rangle=0$ since the ground state is massless
in critical N=2 superconformal field theories. 
Because of the N=2 algebra, these conditions are satisfied if 
\begin{equation}
L_0 |\Phi\rangle = G_{1\over 2}|\Phi\rangle = 
\bar G_{1\over 2}|\Phi\rangle = 
J_0 |\Phi\rangle= J_1 |\Phi\rangle=0, 
\end{equation}
which imply 
that $\Phi$ has no double pole with 
$G$, $\bar G$, $T$ and $J$, and has no single pole with $J$. 

Gauge transformations of $\Phi$ are described by 
\begin{equation}
\delta|\Phi\rangle = 
G_{- {1\over 2}}|\Lambda\rangle  
+\bar G_{- {1\over 2}}|\bar\Lambda\rangle,   \label{27}
\end{equation}
or equivalently, $\delta \Phi = G(\Lambda) +\bar G(\bar\Lambda)$
where $G(\Lambda)$ means to take the contour integral of $G$ around
$\Lambda$. 
These gauge transformations leave invariant
the integrated form of the vertex operator, which is given by
\begin{equation}
\int d^2 z 
G(\bar G(\Phi))
\end{equation}
where $G(\bar G(\Phi))$ means to take the contour integral
of $G$ and $\bar G$ around $\Phi$. 

For the compactification-independent massless states, $\Phi$
must be U(1)-invariant and conformal weight zero, which means it is
a scalar superfield $V(x,\theta,\bar\theta)$. 
No double poles with the N=2 generators implies 
\begin{equation}
(D)^{2}V = (\bar{D})^{2}V = \partial_{m}\partial^{m}V  = 0,
\end{equation}
which are the equations of motion in Lorentz gauge for the
N=1 D=4 super-Maxwell prepotential. The gauge transformations implied
by (\ref{27})
take the usual form 
$\delta V = (D)^2 \lambda + (\bar D)^2 \bar\lambda$ where 
$\Lambda= e^{-i\rho}\lambda$ 
and $\bar\Lambda= e^{i\rho}\bar\lambda$.  

\section{Vertex Operator for First Massive State} 

\noindent

Now we will construct the vertex operator for
the first massive states of the four-dimensional open superstring which
are independent of the compactification.
This N=1 D=4 multiplet of states contains
eight
on-shell bosons and eight on-shell fermions which 
include a massive spin-2 tensor, spin-1 vector, 
and spin-3/2
Rarita-Schwinger field. 

Since the vertex operator represents massive states, it should
contain 
conformal weight one at zero momentum. In other words, the
$e^{-ik^\mu x_\mu}$ dependence contributes conformal weight $-1$,
so the zero-momentum part of the vertex operator should contribute
conformal weight $+1$ in order that the vertex operator
has no second-order pole with the
stress-energy tensor. 

The most general such vertex operator which is U(1)-invariant
and independent of 
the compactification fields is 
\begin{equation}
\Phi = d^{\alpha}W_{\alpha}(x,\theta,\bar\theta) 
+ \bar d^{\dot\alpha}\bar W_{\dot\alpha}(x,\theta,\bar\theta) 
+ \Pi^{m}V_m(x,\theta,\bar\theta)
\end{equation}
\begin{equation}
+\partial\theta^{\alpha}V_{\alpha}(x,\theta,\bar\theta)
+\partial\bar\theta^{\dot\alpha}\bar V_{\dot\alpha}(x,\theta,\bar\theta) 
+\partial\rho V(x,\theta,\bar\theta).
\end{equation}

The 
condition that $\Phi$ is a physical state of the superstring implies 
that $\Phi$ has no double poles with $J$, $G$, $\bar G$, and $T$. 
The vanishing of the double 
poles of the OPE of $J$ with $\Phi$ gives the condition:
\begin{equation}
V = 0.
\end{equation}
The vanishing of the double pole with $G$
gives the following constraint 
equations on the superfields:
\begin{equation}
D^{2}V_{\alpha} = 0 \label{28}
\end{equation}

\begin{equation}
D^{2}{\bar W}_{\dot\alpha} = 0
\end{equation}

\begin{equation}
D^{2}V_{n} 
- 2i(\sigma_{n})_{\alpha \dot{\alpha}}D^{\alpha}{\bar W}^{\dot\alpha} = 0
\end{equation}

\begin{equation}
D^{2}W_{\alpha} 
- 2i(\sigma^{n})_{\alpha \dot{\alpha}}\partial_{n}{\bar W}^{\dot\alpha} 
+ 2V_{\alpha} = 0
\end{equation}

\begin{equation}
D^{2}{\bar V}_{\dot\alpha} 
-2i(\sigma^{n})_{\alpha \dot{\alpha}}D^{\alpha}V_{n} 
+4{\bar W}_{\dot\alpha} = 0, \label{29}
\end{equation}
where the covariant derivatives are defined by
\begin{equation}
D_{\alpha} = \frac{\partial}{\partial\theta^{\alpha}} 
+ \frac{i}{2} 
{\bar\theta}^{\dot\alpha}(\sigma^{n})_{\alpha \dot{\alpha}}\partial_{n}
\end{equation}

\begin{equation}
{\bar D}_{\dot\alpha} = \frac{\partial}{\partial{\bar\theta}^{\dot\alpha}} 
+ \frac{i}{2} \theta^{\alpha}(\sigma^{n})_{\alpha \dot{\alpha}}\partial_{n}.
\end{equation}
The vanishing of the double pole with $\bar G$ gives the complex conjugate
of the above equations. 
Finally, the vanishing of the double pole with $T$ gives the expected
mass-shell condition: 
\begin{equation}
(\partial^{m}\partial_{m} + 2)\Phi = 0. \label{50}
\end{equation}

\noindent

The gauge invariance for the massive state is defined analogously to the 
massless case. 
We define the gauge parameter as
\begin{equation}
\Lambda = e^{-i\rho}\lambda,\quad
\bar\Lambda = e^{i\rho}\bar\lambda,\quad
\end{equation}

\begin{equation}
\lambda = d^{\alpha}C_{\alpha} + \bar{d^{\dot\alpha}}\bar{E_{\dot\alpha}} 
+ \Pi^{m}B_{m} + \partial\rho F
+\partial\theta^{\alpha}B_{\alpha} + 
\partial{\bar\theta}^{\dot\alpha}{\bar H}_{\dot\alpha}.
\end{equation}
and the gauge 
transformation is $\delta\Phi = G(\Lambda) +\bar G(\bar\Lambda).$

It is easy to check that $\delta W^\alpha =2 B^\alpha + ...$ and
$\delta \bar W^{\dot\alpha} =2 \bar B^{\dot\alpha} + ...$ under
this gauge transformation. Therefore, one can choose the gauge
where 
\begin{equation}
W^\alpha = \bar W^{\dot\alpha}=0. 
\end{equation}

In this gauge, the equations (\ref{28})- (\ref{29}) and (\ref{50})
simplify to 
\begin{equation}
V^\alpha = \bar V^{\dot\alpha}= 
W^\alpha = \bar W^{\dot\alpha}= V =0 ,
\end{equation}
\begin{equation}
(\partial^{n}\partial_n +2) V_m=0\label{60}
\end{equation}
\begin{equation}
(\sigma^{n})_{\alpha \dot{\alpha}}D^{\alpha}V_{n} = 0 \label{30}
\end{equation}
\begin{equation}
(\sigma^{n})_{\alpha \dot{\alpha}}\bar{D}^{\dot{\alpha}}V_{n} = 0 \label{31}
\end{equation}

To show that equations (\ref{60}), (\ref{30}) and (\ref{31})
for $V_m$ describe an N=1 D=4 massive spin-2
multiplet, we
expand the superfield $V_{m}$ in components as 
\begin{eqnarray}
V_{m} = C_{m} + i\theta\chi_{m} - i\bar{\theta}\bar{\chi}_{m} \\ \nonumber 
+ i\theta\theta(M_{m} + iN_{m}) - i\bar{\theta}\bar{\theta}(M_{m} - iN_{m}) \\ 
\nonumber
 + \theta\sigma^{n}\bar{\theta}v_{mn} + 
 i\theta\theta\bar{\theta}\bar{\lambda}_{m} 
 - i\bar{\theta}\bar{\theta}\theta\lambda_{m} \\ \nonumber 
 + \theta\theta\bar{\theta}\bar{\theta}(D_{m} 
7- \frac{1}{16}\partial^{n}\partial_{n}C_{m}). 
\end{eqnarray}

In components, equations(\ref{30}) and (\ref{31}) are given by
\begin{equation}
(\sigma^m)_{\alpha\dot{\alpha}}\chi^{\alpha}_{m} = 0 \label{40}
\end{equation}
\begin{equation}
\frac{i}{2}(\bar{\sigma}^{n}\sigma^{m})^{\dot{\rho}}_{\dot{\alpha}}
\partial_{n} C_{m} 
+ (\bar{\sigma}^{n}\sigma^{m})^{\dot{\rho}}_{\dot{\alpha}}v_{mn} = 0 \label{32}
\end{equation}
\begin{equation}
(\sigma^m)_{\alpha\dot{\alpha}}(M_{m} + iN_{m}) = 0  \label{33}
\end{equation}
\begin{equation}
\frac{1}{4}(\bar{\sigma}^{n}\sigma^{m})^{\dot{\rho}}_{\dot{\alpha}}
\partial_n \bar{\chi}_{m \dot{\rho}}
- i(\sigma^{m})_{\alpha {\dot{\alpha}}}\lambda^{\alpha}_{m} = 0  \label{34}
\end{equation}
\begin{equation}
(\bar{\sigma}^{n}\sigma^{m})^{\dot{\rho}}_{\dot{\alpha}}
\partial_n (M_{m} + iN_{m}) = 0
\end{equation}
\begin{equation}
\frac{-i}{4}(\sigma^{n}\bar{\sigma}^{p}\sigma^{m})_{\alpha\dot{\alpha}}
\partial_{p}v_{mn}
 + 2(\sigma^m)_{\alpha\dot{\alpha}}(D_{m} - \frac{1}{16}
 \partial^{n}\partial_{n}C_{m}) = 0  \label{35}
 \end{equation}
 \begin{equation}
(\bar{\sigma}^{n}\sigma^{m})^{\dot{\rho}}_{\dot{\alpha}}
\partial_{n}\bar{\lambda}_{m \dot{\rho}} = 0 .  \label{36}
\end{equation}
 
We now analyse the propagating degrees of freedom of the superfield $V_m$. 
Firstly, we analyse the bosonic sector. Equation (\ref{33}) implies 
$M_m = N_m = 0$. Taking the trace of the equation (\ref{32}), we get the 
conditions $\partial^m C_m = v_m{}^m = 0$. Contracting equation (\ref{32}) 
with $(\bar{\sigma}^{p}\sigma^{q})^{\dot{\alpha}}_{\dot{\rho}}$, we 
obtain a relation between the antisymmetric part of $v_{mn}$ and $C_m$, 
namely
\begin{equation}
v_{[mn]} = \frac{1}{4}\epsilon_{mnpq}\partial^{p}C^{q}. \label{37}
\end{equation}
Equations (\ref{35}) implies 
$\partial^{m}v_{(mn)} = 0$ and, together with 
(\ref{37}), implies $D_m = 0$. So the bosonic fields in $V_m$ describe
a massive vector $C_m$ and a massive symmetric traceless tensor
$v_{(mn)}$ satisfying the Lorentz gauge conditions 
$\partial^{m}C_m  $ = 
$\partial^{m}v_{(mn)} = 0.$

For the fermionic sector, 
equations (\ref{40}) and (\ref{34}) imply
\begin{equation}
\partial^m \bar\chi_{m \dot\alpha} = 
\partial^m \chi_{m \alpha} = 0,\label{45}
\end{equation}
\begin{equation} 
(\sigma^{m})_{\alpha\dot{\alpha}}\chi_{m}^{\alpha} = 
(\sigma^{m})_{\alpha\dot{\alpha}}
\bar{\chi}_{m}^{\dot\alpha} = 0, \label{42}
\end{equation}
and 
\begin{equation}
(\sigma^{n})_{\alpha\dot{\alpha}}\partial_{n}{\bar\chi}_{m}^{\dot{\alpha}} = 
4i \lambda_{m}^{\alpha} \;,\;\;\;\;
(\bar{\sigma^{n}})^{\dot{\alpha}\alpha}\partial_{n}\chi_{m \alpha} =- 4i  
{\bar\lambda}_{m}^{\dot{\alpha}}. \label{43}
\end{equation}
Equations (\ref{45})-(\ref{43}) and (\ref{60})
describe a massive spin-3/2 Dirac
fermion, 
\begin{equation}
\Psi_m^A =~( ~\chi_{m}^\alpha,~  
2\sqrt{2} ~\bar\lambda_m^{\dot\alpha}~),
\end{equation}
satisfying
the Lorentz gauge condition $\partial^m \Psi_m^A$ =0.  

The superfield $V_m$ therefore provide a superspace
representation of the 8+8 on-shell 
degrees of freedom of the N=1 D=4 massive spin-2 multiplet. 

\section {Acknowledgements}

NB would like to acknowledge partial support from CNPq grant number
300256/94-9 and MML would like to acknowledge support from FAPESP
grant number 96/03546-7.

\end{document}